\documentclass[floatfix,%
 reprint,
 amsmath,amssymb,
 aps,
]{revtex4-2}

\usepackage{graphicx}
\usepackage{dcolumn}
\usepackage[colorinlistoftodos]{todonotes}
\usepackage{bm}
\usepackage{ulem}
\usepackage{xcolor} 

\usepackage{tabularx}
\usepackage{multirow}
\usepackage{appendix}
\usepackage{array}
\usepackage{booktabs}   
\usepackage{enumitem}   

\begin{document}

\preprint{APS/123-QED}

\title{Trajectories into Careers in the Quantum Industry: Beyond Knowledge and Skills}

\author{Shams El-Adawy\textsuperscript{1,2}}

\author{A.R. Pi\~na\textsuperscript{3}}

\author{Benjamin M. Zwickl\textsuperscript{3}}

\author{H. J. Lewandowski\textsuperscript{1,2}}

\affiliation{\textsuperscript{1}JILA, National Institute of Standards and Technology and the University of Colorado, Boulder, Colorado 80309, USA}
\affiliation{\textsuperscript{2}Department of Physics, University of Colorado, Boulder, Colorado 80309, USA }
\affiliation{\textsuperscript{3}School of Physics and Astronomy, Rochester Institute of Technology, Rochester, NY 14623, USA}

\date{\today}

\begin{abstract}
Career preparation for participation in the quantum industry is often framed in terms of formal educational pipelines, workforce projections, and knowledge and skills needed for various roles. Less is known about how quantum industry professionals themselves characterize their preparation for participation in the field. In this paper, we analyze interviews with quantum industry professionals working across a range of positions and company types to examine the experiences that enable entry into the quantum industry. Using thematic analysis, we identify four trajectories: (1) continuity of research practice from academia to industry, (2) reframing of prior expertise for quantum applications, (3) incremental engagement through various professional opportunities, and (4)  network-enabled entry. These trajectories often co-occur within individual narratives, showing that preparation emerges from a combination of educational, professional, and relational experiences. Our findings demonstrate that supporting preparation for quantum industry careers requires more than the design of formal coursework and degree programs. Our results highlight the importance of experiential learning opportunities that allow students to apply their knowledge and skills and develop professional connections that facilitate entry into quantum careers.

\end{abstract}

\maketitle


\section{Introduction}

Quantum Information Science and Engineering (QISE) has rapidly developed into an interdisciplinary field with cutting-edge research and development being conducted in universities, national laboratories, startups, and large technology companies. As advances in quantum computing, sensing, networking, and communication continue, there has been growing attention to the development of a future quantum workforce \cite{NQI, NSP, CHIPS, nstc_qist_workforce_plan_2022, taylor2023us}. In particular, researchers have begun to examine how educational programs can support students' preparation for quantum careers \cite{fox2020preparing,kaur2022defining,greinert2023future, greinert2024advancing, li2024developing, rosenberg2024science, el2025insights, oliver2025education}.  

Current research on quantum career preparation has largely emphasized the development of educational pipelines into QISE \cite{meyer2024disparities, el2025insights,pina2025landscape, pina2025investigating, fox2020preparing, greinert2024advancing, oliver2025education}, the identification of knowledge, skills, and abilities (KSAs) needed for existing quantum-related jobs \cite{hughes2022assessing, Pina2025QuantumWorkforce, el2026experimental}, and how education systems might align with anticipated growth in the field \cite{greinert2023future, el2025industry, devendrababu2025mapping, el2025coloradoreport}. One of the main goals of this research is to characterize what knowledge and skills are valued for positions in the quantum industry in order to guide course and program development. 

Much of this research frames preparation for participation in the quantum industry in terms of formal training and credential attainment. Within this framing, entry into the quantum industry is often simplified as a process in which acquiring disciplinary knowledge and skills directly leads to workforce readiness. While this perspective is useful for guiding course and program development, it does not fully capture how individuals actually become able to participate in the quantum industry. In fact, the broader STEM education literature has shown that linear pipeline models oversimplify how individuals move into and across scientific and technical fields by giving insufficient attention to the role of social and cultural capital that lead to the nonlinear pathways through which individuals enter STEM careers \cite{2014problematizingSTEMpipeline}. This limitation is particularly salient in emerging and interdisciplinary fields such as QISE, where roles, expectations, and career pathways into the field are still evolving alongside the field itself \cite{Forbes2010}. Thus, focusing primarily on knowledge and skills for course and program development may miss the broader range of experiences that individuals navigate, combine, and interpret in order to enter the field.

Drawing on a situated learning perspective \cite{lave1991situated} can offer a useful way to rethink this limitation by conceptualizing learning as increasing participation in the socially organized practices of a community rather than solely the acquisition of knowledge and skills.  Becoming prepared for a field involves more than completing coursework or earning a degree. It also includes gaining access to authentic activities, developing familiarity with how work is done in the field, and being recognized by others as able to contribute meaningfully \cite{wenger1999communities, wenger2014learning,gee2000}. Prior research has shown that learning unfolds through engagement across multiple contexts and over time, and is shaped through the accumulation and interpretation of these various experiences \cite{wenger1999communities, wenger2014learning}.

This perspective suggests that preparation for participation in QISE should be understood as more than simply acquiring knowledge and skills, but as a process through which individuals assemble and interpret a range of experiences that position them to enter the field.  Hence, examining how quantum industry professionals themselves describe their trajectories into QISE can provide insight into the experiences and forms of engagement that enable entry into this rapidly evolving field. Such insight can inform physics educators who are designing and developing curricula, experiential learning opportunities, and professional development structures that support students' preparation for quantum careers.   

Therefore, in this study, we shift the focus from identifying required knowledge and skills to examining how individuals are enabled to enter QISE careers and contribute to QISE broadly.  Specifically, we address the following research question:
\textbf{How do quantum industry professionals describe their trajectories into the quantum industry?}

\textit{Note on terminology:} In this paper, we use \textit{QISE} to refer to the broader interdisciplinary area of quantum information science and engineering, which spans research and development in academia and industry. We use \textit{quantum industry} more specifically to describe employment contexts in which quantum technologies are developed and applied, including startups and larger established companies \cite{QEDC2026}. We also use terms such as \textit{field} and \textit{community} to refer broadly to the interconnected professional spaces, networks, and activities through which individuals gain exposure and engage with quantum-related work, develop relevant experience, and access opportunities. We do not treat these terms as referring to a single bounded community. Instead, consistent with our focus on trajectories, participation in QISE is understood as occurring across multiple overlapping contexts that individuals navigate as they become positioned to enter the quantum industry. Lastly, by \textit{trajectories}, we refer to the experiences and factors that shape the pathways by which quantum industry professionals enter the quantum industry.

\section{Background}\label{Background}

\subsection{Quantum education and workforce development}
Research on quantum education and workforce development has primarily focused on the design of QISE courses and programs and the identification of knowledge, skills, and abilities (KSAs) relevant for quantum industry roles to best prepare students for careers in this field.

A substantial body of work examines how educational systems can support entry into quantum careers. Researchers have examined how higher education provides routes
into the quantum industry through degrees in physics, engineering, and computer science \cite{meyer2022today, meyer2024introductory, pina2025landscape}. Other research has focused on the design of specific courses that provide quantum knowledge and skills to prepare students to enter the workforce directly after their
undergraduate studies \cite{porter2022creating,meyer2024introductory, oliver2025education}. These studies often frame preparation as progression through sequences of courses that build foundational and specialized knowledge aligned with the needs of the quantum industry.

In parallel, several studies have sought to identify the specific knowledge, skills, and abilities (KSAs) relevant to quantum industry roles. Results from this research strand characterize the combination of technical expertise and professional skills valued across quantum-related roles \cite{fox2020preparing, hughes2022assessing}. In particular, researchers have explored how quantum-specific knowledge and experimental skills could inform course and program design \cite{greinert2023future, Pina2025QuantumWorkforce, el2026experimental}. Additionally, researchers have investigated the scale of participation required to support the growth of quantum technologies. For instance, some studies have underscored how investment across all educational levels, spanning K-12 and higher education, is needed to align with anticipated growth of QISE \cite{asfaw2022building, greinert2023future, greinert2024advancing, el2025industry, devendrababu2025mapping, goorney2026quantum,ushousequantum,emilyedwards2026introducing}.

Across this body of work, preparation is frequently conceptualized in terms of formal training and credential attainment, including degrees, specialized coursework, and skill development aligned with identified industry-valued KSAs. While this research is useful for understanding how preparation may be structured, it provides limited insight into how individuals navigate their career pathways into the quantum industry in practice. In particular, prior research may under report the roles of experiential learning, professional development, and professional networks that also shape entry into the field. 

\subsection{Research on career preparation}

Research on STEM career preparedness offers a broader perspective on how individuals become ready to enter professional roles, emphasizing that preparation extends beyond formal coursework or degree completion. The literature in this space highlights the importance of experiential and social processes that unfold across multiple contexts. 

Studies of doctoral and early career scientists show that professional development experiences play a critical role in shaping career trajectories. For example, Ganapati \textit{et al.} identify internships, research collaborations, and exposure to various career pathways as key components of career readiness \cite{ganapati2021professional}. Similarly, Edwards \textit{et al.} demonstrate that career pathways are often nonlinear, involving transitions across sectors, roles, and areas of expertise rather than following a single trajectory \cite{edwards2023mapping}.

Within physics education research (PER), similar patterns have been investigated, with studies emphasizing that preparation is mediated through engagement in activities and communities that extend beyond formal instruction. For example, participation in informal physics programs has been shown to increase the interest and relevance of physics and
science as potential career paths for individuals engaged in them \cite{fracchiolla2020community,bell2020informal, prefontaine2021informal, el2024motivation}. These experiences outside of formal instruction provide opportunities for students to encounter authentic practices that are part of professional physics practice.

Moreover, research on undergraduate and early research experiences further illustrates how preparation is shaped through participation in authentic professional activities. For example, undergraduate research experiences have been shown to influence students’ career decisions and orientation toward a particular subfield, while also fostering relationships and communities that support further opportunities \cite{rosa2016educational, NASEM2017, zohrabi2022impact}. Additionally, course-based undergraduate research experiences have been shown to positively shape students’ trajectories into physics \cite{werth2022impacts, oliver2023student}. Across these different types of research experiences, preparation for the next career step has been shown to occur through engagement in activities that approximate professional practice and through interactions with members of the field. 

Research on career preparation conceptualizes workforce development as experiential, socially mediated, and unfolding across contexts and over time. However, while this literature establishes the importance of these processes, it does not show how such experiences collectively enable entry into an evolving field such as QISE. In particular, it remains an open question how individuals interpret and assemble these experiences in ways that enable access to opportunities within the quantum industry.  

\subsection{Situated learning perspective}
 To better understand how various experiences across formal education, professional development opportunities, and networks collectively enable entry into the field, we draw on a situated learning perspective, which provides a broader lens for understanding preparation as a process of gaining access to participation in a professional community \cite{lave1991situated}. This perspective aligns with Sfard's distinction between learning as the acquisition of knowledge and skills and learning as participation in a community, arguing that both offer complementary insights into learning, while emphasizing participation as particularly useful for understanding how individuals prepare to enter professional communities \cite{sfard1998two}. 
 
 A situated learning perspective provides a way to understand preparation not simply as the acquisition of knowledge or skills, but as a trajectory of increasing participation in socially organized practice \cite{lave1991situated, wenger1999communities}. Learning is understood as becoming able to engage in the practices of a community through participation in different experiences over time. Preparation therefore involves gaining access to authentic activities, developing familiarity with how work is carried out in a field, and being recognized by others as a legitimate contributor within that field. From this view, entry into a professional field is mediated not solely by what individuals know, but by how combinations of experiences enable them to engage in, and be recognized within, relevant contexts. 

 This perspective is particularly relevant for QISE, where pathways are not standardized and participation often involves navigating an interdisciplinary context across academia and industry. Rather than assuming a linear progression from education to employment, a situated learning lens allows us to examine how individuals build and interpret experiences such as engagement with research and professional connections that enable entry into the quantum industry. 

Although past PER studies using a situated learning lens have primarily focused on contexts such as upper-division physics laboratory
courses and informal physics programs \cite{irving2015becoming, rethman2021impact}, they similarly emphasize the key role of participation in authentic shared practices and the importance of recognition in shaping access to opportunities. Extending this perspective to the context of QISE education research enables us to analyze how individuals become positioned for participation in an emerging field. 

Thus, trajectories into the quantum industry can be understood as pathways through which individuals gain access to relevant practices, reinterpret prior experiences as legitimate forms of participation, and are recognized by others as able to contribute to the professional community.  This framing guides our analysis of how quantum industry professionals characterize the experiences that enabled entry into QISE careers.

\section{Methods} \label{Methods}

\subsection{Data collection}
This study draws on a large corpus of research interviews with professionals at quantum industry positions based in the United States. In brief, interview participants were recruited using a combination of purposeful and convenience sampling \cite{etikan2016comparison}, including outreach through the research team's professional networks, and snowball sampling \cite{parker2019snowball} within companies (see \cite{el2026experimental} for additional methodological details on recruitment).

The full dataset consists of 75 interviews conducted between December 2024 and July 2026 with employees (N=36) and managers (N=39) across 33 different quantum companies, representing a range of company types and activities. Interviews were conducted over Zoom, lasted approximately one hour, and followed one of two interview protocols. One interview protocol was designed for managers to elicit organization-level perspectives on roles and required knowledge, skills, and abilities (KSAs), whereas the other interview protocol was for employees to capture task-level experiences and individual career trajectories. Full interview protocols can be found in \cite{el2026experimental}.

\begin{figure}
    \centering
    \includegraphics[width=1.0\linewidth]{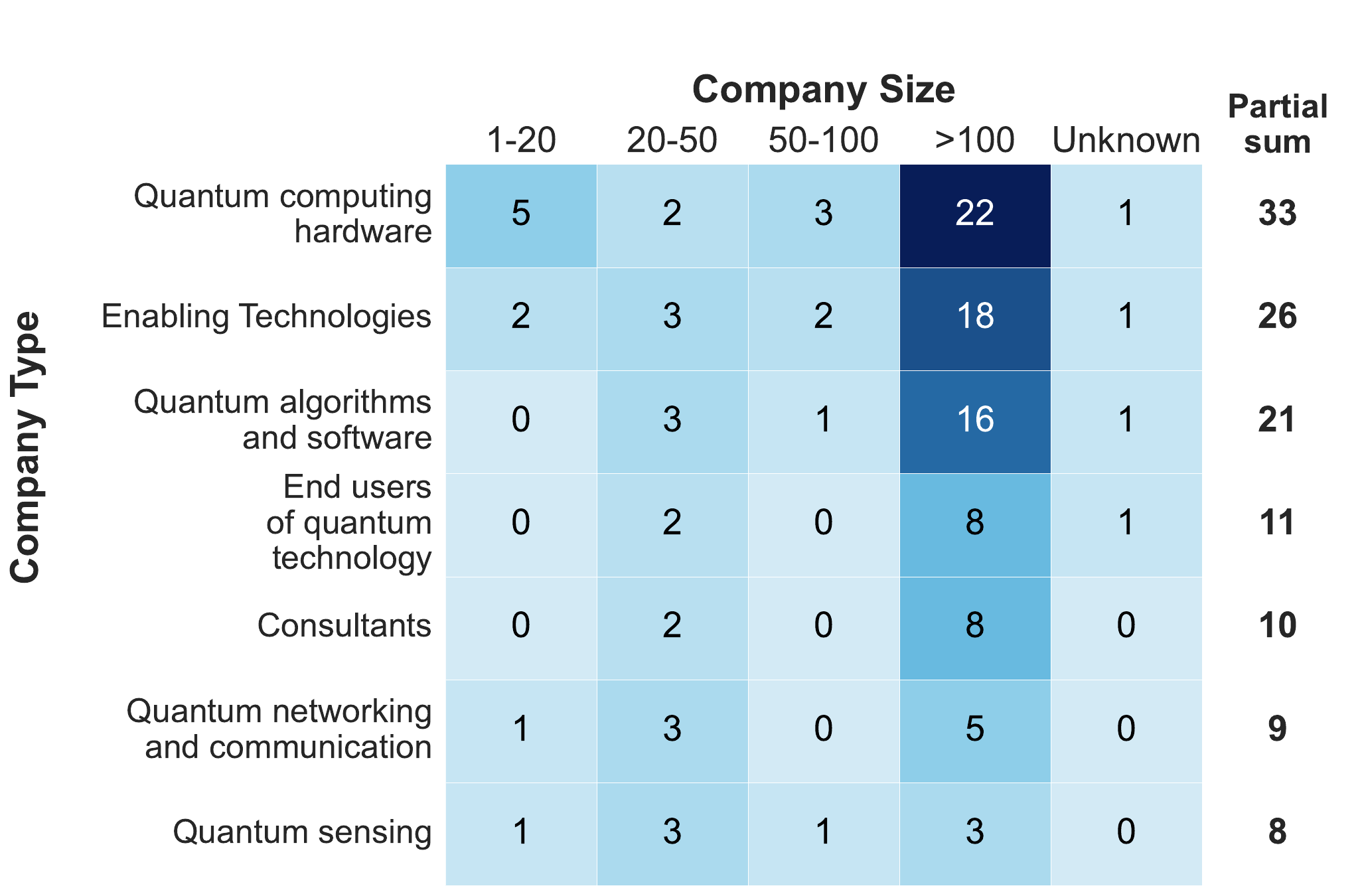}
    \caption{Distribution of the 36 employee interviews across company type and size. Companies are categorized by activity type following definitions in prior literature \cite{el2025industry}. Company size corresponds to the interviewee-reported range of employees working in quantum-related technologies at their company. The numbers in each cell indicate the number of interviews associated with a given company type (rows) and size category (columns). The dataset includes multiple interview participants from the same company, and because companies may be classified into multiple activity types, a single interview may contribute to more than one company type. Row-wise partial sums show the total number of interviews associated with each company type across all sizes. }
\label{fig:companycharacterization}
\end{figure}

The analysis presented in this paper focuses specifically on the subset of employee interviews (N=36) who represent 21 distinct companies, as these interview participants were directly asked about their educational and career pathways. In particular, we analyze responses to questions related to educational background and career preparation in the employee interviews, including (i) their academic training (degree and disciplinary backgrounds), (ii) prior work experience expected for entry into their current roles, (iii) professional development experiences during their academic training (e.g., internships, research projects, workshops, and conferences) that they perceive as relevant to their current work, and (iv) coursework or content knowledge that they identify as most applicable to their current positions. 

To contextualize the subset of employee interviews analyzed in this study, Figure \ref{fig:companycharacterization} summarizes the interview participants' company types and sizes, while the left-hand side of Figure \ref{fig:trajectories_co-occurrence} summarizes their roles and educational backgrounds. The sample includes individuals in a range of positions spanning all position types (hardware, software, bridging, and public-facing and business roles, following the categorization in Ref. \cite{pina2025categorization, el2026profiles}), and reflects various educational levels and disciplinary backgrounds, including physics, math, engineering, computer science, and other related disciplines, reflecting the interdisciplinary nature of QISE. 

\subsection{Data analysis}
We conducted a thematic analysis \cite{braun2006using} to characterize how interview participants described their preparation to engage in the quantum industry. Our analysis focused on how interviewees constructed and made sense of their preparation for entering and contributing to the field. Rather than treating preparation as a fixed sequence of educational steps, we examined how participants connected and interpreted their experiences as enabling participation in their current roles.

The unit of analysis consisted of excerpts within interviews in which participants described aspects of their preparation, including educational experiences, research involvement, professional development activities, and transitions into industry roles.  Initial coding involved identifying segments within these excerpts in which participants described experiences they framed as enabling their entry into the quantum industry. These segments were compared across interviews to identify similarities and differences in how preparation was framed. Through an iterative process, we examined how participants connected multiple experiences over time, focusing on how these experiences were assembled and interpreted as enabling entry into QISE careers. Recurring patterns across participants were grouped into themes that captured distinct ways of narrating preparation.

We used the term \textit{trajectories} to refer to these themes. These trajectories referred to recurring ways through which interview participants described gaining access to relevant practices, reframing prior experiences as legitimate participation, and becoming recognized as able to contribute to the quantum industry.  These trajectories did not represent fixed or linear career pathways, but rather characterizations of how individuals constructed and interpreted their readiness for participation. Moreover, these trajectories  were not mutually exclusive, i.e., individual interview participants often drew on multiple trajectories within a single narrative. These simultaneous or sequential occurrences reflected how multiple forms of participation across contexts collectively contributed to enabling entry into the field.

Coding was conducted collaboratively between the first and second authors of this paper. SE conducted initial coding across all employee interviews to identify excerpts in which participants described experiences that enabled entry into the quantum industry. Preliminary codes captured both the type of experience described (e.g., research and internship) and the function of that experience within the interviewee's narrative (e.g., enabling application of knowledge and providing access to a quantum-specific network). These preliminary codes were then grouped into broader trajectory categories, which were refined iteratively. 

To establish  inter-rater reliability (IRR), AP independently applied the trajectory definitions to all interviews. Agreement was evaluated at the level of trajectory identification within each interview (i.e., whether a given trajectory was present or not within a participant's narrative). The IRR process was conducted across three rounds, each on a distinct subset of interviews including approximately one third of the interviews each time, such that all 36 employee interviews were independently coded by both researchers by the end of the process. Pre-discussion agreement improved from 60\% in round one, to 67\% in round two, and 82\% in round three. For each round, discrepancies were resolved through discussion, resulting in full consensus on the trajectories present in each interview. 

These discussions about discrepancies between coders involved revisiting the full interview context to ensure that trajectory categorization reflected participants' broader narratives rather than isolated excerpts and clarifying trajectory definitions. For example, we refined Trajectory 1 (T1) to include research experiences relevant to participants' current industry position, even when those experiences were not explicitly related to QISE. Also, Trajectory 2 (T2) was refined to explicitly capture prior work experience outside of quantum-specific roles. This process resulted in the final articulation of the four trajectories summarized in Table \ref{tab:trajectoriessummary}. 

\subsection{Limitations}
A couple of limitations are important to note for the interpretation of our results. First, although our dataset is large, the number of interviewees limits the generalization of our findings. The identified trajectories should be interpreted as illustrative rather than exhaustive of all possible career pathways into the quantum industry. Second, the interview protocols were originally designed to elicit knowledge and skills relevant to ongoing quantum industry work. As a result, some aspects of past career development may not have been explored in depth. Participants’ discussions of preparation therefore reflect the experiences they chose to highlight within the limited set of questions that elicited reflections on career pathways.

\section{Results}\label{Results}

Across interviews, participants narrated their entry into the quantum industry by connecting prior experiences to their ability to contribute to current roles. We conceptualized these patterns as trajectories of participation, which are summarized in Table \ref{tab:trajectoriessummary}. These trajectories captured distinct, yet simultaneous and/or sequential, ways in which participants constructed their preparation by linking past experiences to their ability to contribute within the quantum industry. 

Before we provide an overview of the trajectories and illustrative cases, it is important to note that all interview participants described their educational background in terms of formal degrees and disciplinary training as indicated on the left-hand side of Figure \ref{fig:trajectories_co-occurrence}, most commonly in physics, math, engineering, or computer science. This disciplinary training established the technical foundation from which interviewees described their preparation. However, participants did not frame degree attainment or coursework alone as the only preparation needed for entry into the quantum industry. In particular, interviewees discussed preparation through combinations of experiences that enabled them to apply that knowledge, engage in relevant practices, and access opportunities through which their participation in the field became possible.

\begin{table*}[htbp]
\caption{Summary of the trajectories describing preparation for participation in the quantum industry}
\label{tab:trajectoriessummary}
\begin{ruledtabular}
\begin{tabular}{ll}
\textbf{Trajectory} & \textbf{Brief description}
\\ \hline
T1: Academic research aligned with industry needs & Continuity of research practice between academic research and industry \\ & work
\\
T2: Reframed technical background
& Reframing expertise from prior technical work for quantum applications
\\
T3: Incremental engagement
& Gradually increased participation in quantum-related activities \\ &  (e.g., summer schools, workshops, conferences, internships)
\\
T4: Network-enabled entry
& Entry facilitated by existing members of the quantum industry \\

\end{tabular}
\end{ruledtabular}
\end{table*}

\subsection{Overview of trajectories}

\subsubsection{Trajectory 1: Academic research training to industry alignment}

In trajectory one (T1), preparation is framed as continuity of practice between academic research and industry work. Interview participants describe their prior research experience, most often at the doctoral level, as already constituting the type of work performed in their current roles, such that entry into industry is framed as a shift in context from an academic to a commercial setting rather than a change in practice. 

Of the 36 interviewees, 25  described alignment between their academic research and their current industry position. Variation within this trajectory reflected different levels of alignment. For some participants, alignment occurred at the level of specific technological systems, where their PhD research involved designing or operating the same types of quantum hardware used in their companies (e.g., conducting PhD research on trapped-ion quantum computing and continuing to do research on the same hardware in industry). Others described alignment at the level of technical subfield, where experimental research experience provided familiarity with practices, tools, and problem-solving approaches that were recognizable in quantum-related industry work (e.g. their doctoral research in the subfield of atomic, molecular, and optical physics was used in their current industry role). In this trajectory, interview participants emphasized that their research training, whether explicitly in QISE or not, equipped them not only with domain knowledge, but with sustained participation in the practices and tools central to their current roles.

This trajectory highlighted that preparation was not solely about acquiring relevant knowledge, skills, and abilities (KSAs), but about participating in forms of work that are already recognizable as industry-relevant. Hence, academic research experience can function as a site where individuals engage in authentic practices that translate directly to quantum industry work. This trajectory was distinct from subsequent ones we found in that positioning for participation was enabled through continuity of practice, rather than through transferring expertise from prior work experience, gradual entry into the field, or access mediated by professional networks.

\subsubsection{Trajectory 2: Reframed technical background}
In trajectory two (T2), preparation was characterized by the reframing of expertise developed through prior technical work experience. Interview participants described becoming positioned to participate in the quantum industry by demonstrating how prior knowledge and skills developed in other professional contexts can be translated and reinterpreted into quantum-related work.  Unlike T1, where participants discussed continuity between academic research practices and industry work, readiness in T2 was established through the reinterpretation of prior technical expertise as relevant to quantum applications.

Of the 36 interviewees, 17 described entering the quantum industry through transferable expertise from previous work experience. Variation within this trajectory reflected differences in how participants connected prior experiences to quantum technologies. For some, this transition involved having worked in closely related domains such as software engineering, where existing knowledge could be applied to the development of quantum technologies. For others, preparation was framed more broadly in terms of transferable analytical skills, such as computational modeling or mathematical reasoning, that could be applied to new contexts such as QISE. Although participants sometimes referenced coursework in math, physics, computer science, and engineering as a foundation, the defining feature of T2 was the translation of expertise developed through prior work experience into quantum-related contexts.  

This trajectory foregrounded preparation as a process of reframing and translating existing technical expertise to a new context. This trajectory differed from T1 in that prior experiences do not involve direct continuity with quantum-related research practices, but instead reflect the transfer and reinterpretation of expertise from other technical domains. It also differed from subsequent trajectories, as it does not emerge through progressive engagement with quantum-specific activities and/or professional networks.

\subsubsection{Trajectory 3: Incremental engagement}

In trajectory three (T3), preparation emerged as a process of increasing  participation in quantum-related activities over time. Entry into the quantum industry was described as a sequence of engagements that gradually increased participants' familiarity with the field and its practices. Through these engagements, individuals gained increasing access to participate by moving from peripheral exposure toward more central members of the field.  

Of the 36 interviewees, 15 described incremental engagement as part of their preparation, although no participant described this trajectory in isolation. Interviewees emphasized how different forms of engagement contributed to their preparation. Some describe structured programs such as summer schools, internships, or workshops that provided exposure to ongoing work in the quantum community. Others described more informal participation, such as attending QISE talks at conferences or engaging in quantum-related extracurricular activities, which helped them better understand the field. Across these interviewees, incremental engagement functioned as a mechanism for progressively participating in QISE.

This trajectory highlighted preparation as a process of gradual engagement with the field, through which individuals become increasingly familiar with QISE.  Unlike T2, where positioning depended on reframing prior expertise, this trajectory emphasized progressive participation within quantum-specific contexts. It also differed from the subsequent trajectory in that access emerges through accumulated engagement rather than mediated by existing members of the quantum industry.  

\subsubsection{Trajectory 4: Network-enabled entry}
In trajectory four (T4), preparation was enabled through access mediated by recognition from individuals already within the quantum industry. Participants described moments in which connections such as former colleagues, mentors, professional connections, or friends played a decisive role in enabling entry into the industry. While participants may had the relevant KSAs,  entry occurred when those KSAs were recognized and validated within their professional network. 

Of the 36 interviewees, 15 described network-enabled entry. While these participants had relevant skills and knowledge, the critical turning point in their narratives involved recognition or introduction by individuals already working in the quantum industry. Variations within this trajectory reflected different forms of relational connections. Some participants described direct referrals from colleagues or friends, while others described prior professional connections with the organization that later led to employment opportunities. In other cases, familiarity with members of a technical subfield  supported access to relevant job roles. Across interviewees with this trajectory, entry into the quantum industry was facilitated not solely through technical preparation, but through recognition by existing  members of the field. 

This trajectory highlighted that preparation was not solely an individual process, but a socially mediated one. This trajectory was distinct from the others in that preparation was not only an individual process of developing or demonstrating capability, but a relational process in which access depended on existing members of the quantum industry facilitating entry.

\subsubsection{Co-occurrence of trajectories}
\begin{figure*}
    \centering
    \includegraphics[
        width=\textwidth,
        height=\textheight,
        keepaspectratio
    ]{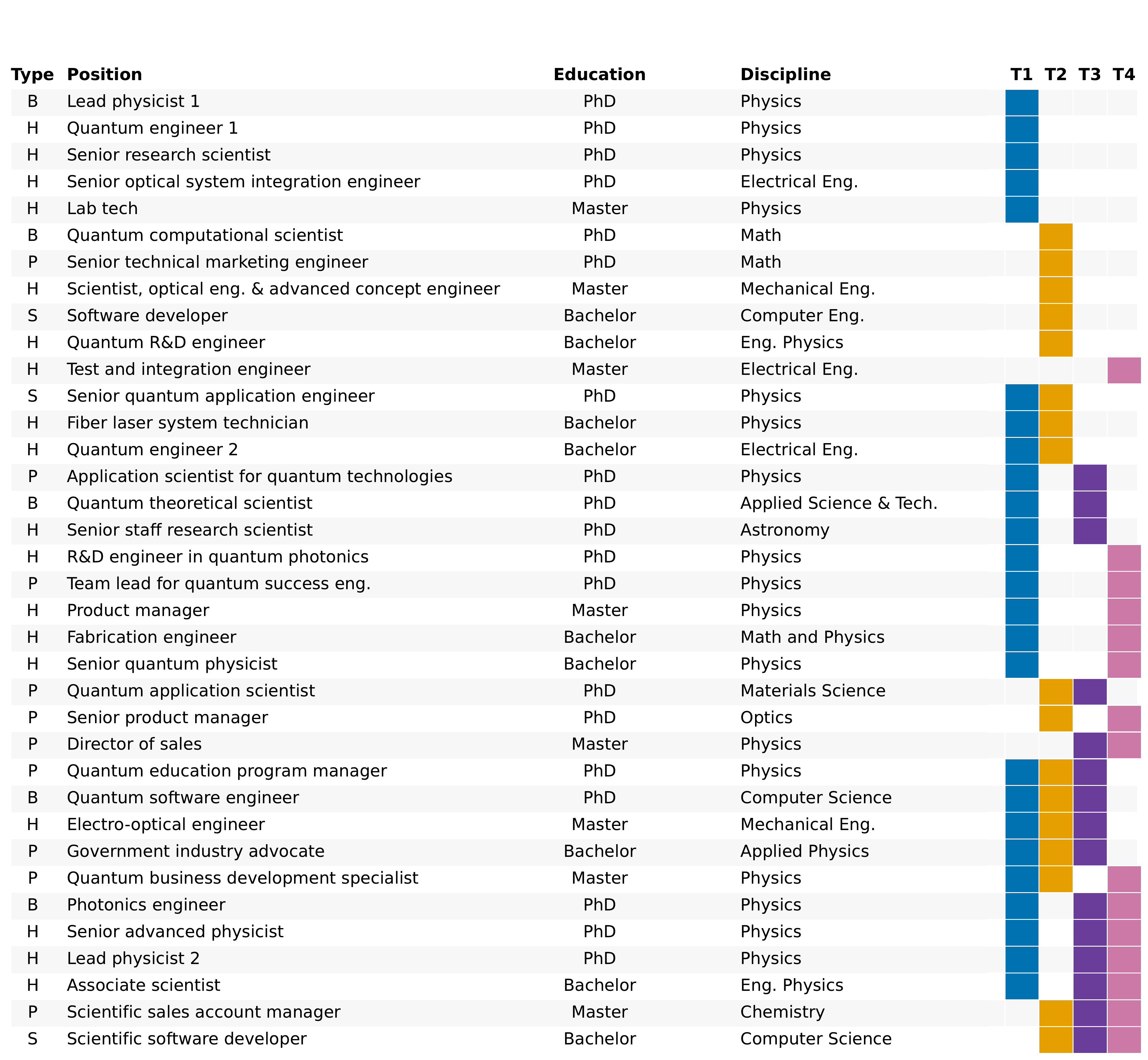}
    \caption{Titles of individual positions, education, discipline, and trajectories (T1, T2, T3, and T4) of quantum industry employees who were interview participants for this study. Twenty-five participants have co-occurring trajectories. The \textit{Type} column indicates role categories: H = hardware roles, B = bridging roles, S = software roles, and P = public-facing or business roles, following the categorization in Ref.~\cite{pina2025categorization,el2026profiles}.}
    \label{fig:trajectories_co-occurrence}
\end{figure*}

Although the trajectories characterized the different types of experiences, interview participants often drew on multiple trajectories within their narratives.  As shown in Figure \ref{fig:trajectories_co-occurrence}, 25 of the 36 interviewees in our dataset leveraged more than one trajectory. Participants often described how different experiences such as research, transferable expertise, incremental engagement, and professional networks work together to enable access to opportunities in the quantum industry. These sequential or simultaneous experiences showed that preparation did not happen through a single pathway, but through the combination of multiple forms of participation across contexts. 

This pattern is consistent with prior research \cite{irving2015becoming, zohrabi2022impact, ganapati2021professional, edwards2023mapping}, which emphasizes that becoming able to participate involve both engagement in relevant practices and access to contexts in which that participation is recognized and valued. In this sense, these trajectories function as complementary mechanisms that collectively contribute to preparation for quantum industry work.

\subsection{Illustrative cases}
To illustrate how the identified trajectories co-occur within individual narratives, we  present four interviewees whose trajectories highlight different configurations. These illustrative cases are selected to highlight variation in how interviewees discussed their preparation.

\subsubsection{Trajectory of a senior optical system integration engineer}

One of our interviewees is currently a senior optical system integration engineer at a company that works on quantum computing hardware, algorithms and software, and networking and communication. Their trajectory into the quantum industry aligns with trajectory one (T1), in which preparation for participation in the company is constructed as direct continuity of research practice: 
\begin{quote}
    My [dissertation] research topic was directly related to my current role. My research was to design, build, test a quantum computer, which is pretty much exactly the same thing I'm doing here in this company.
\end{quote}

This quote illustrates how this participant positions their industry role as a direct continuation of doctoral research practice. Thus, entry into industry is framed not as a transition requiring new forms of preparation, but as a shift in context, from academic research to commercialization of the same technological system.  

The interviewee further emphasizes the continuity of specific experimental skills developed during graduate research training:
\begin{quote}
   My academic research skills, like optics, lasers, vacuum chamber, cryostat, then in a trapped-ion or a trapped neutral atom, then collect the data and write the paper are all very important skills for this position.
\end{quote}

Here, specific experimental skills developed during graduate research are identified as directly used in their current industry work, reinforcing the role of academic research training as valued preparation for industry.

For this senior optical system integration engineer, formal coursework supported the research specialization that later aligned directly with their industry position. The interviewee references engaging in relevant courses as part of the preparation, but emphasizes that research participation, not coursework alone, was the experience most directly aligned with their current position:
\begin{quote}
    In grad school [...]I took a variety of quantum information classes during the PhD program, plus some engineering regular classes. My main focus of the research was AMO physics, [...]which is pretty much the same thing I'm doing here in this company.
\end{quote}

 This quote illustrates how degree attainment and coursework were described as supporting this specialization, but not sufficient on its own. The central preparation occurs through sustained engagement in research practices, which later became recognizable as industry relevant expertise.

\subsubsection{Trajectory of a quantum computational scientist}
One of our interviewees is a quantum computational scientist at a company that wants to use quantum computers to enhance their business. Their trajectory into the quantum industry mostly aligns with trajectory two (T2), in which preparation for participation in the quantum industry is framed through the transferability of expertise from previous work experience. Rather than position their doctoral research as directly aligned with quantum technologies, the interviewee constructs legitimacy through transferable analytical and computational skills:
\begin{quote}
     A lot of what I'm transferring now to my work, looking at quantum algorithms, would be skills in linear algebra, skills in statistics.
\end{quote}
This interviewee, who has a PhD in math does not describe quantum-specific training, but instead identifies mathematical and statistical expertise as transferable to quantum algorithms. This reflects T2, where preparation is established not through prior academic research experience in QISE, but through the reinterpretation of existing expertise. In particular, the interviewee describes work experience involving numerical modeling and large-scale computational problems, which later became relevant to quantum algorithms. The transition into quantum-related work is described as emerging within the organizational context:
\begin{quote}
    We [my team] said, ``Computing in general is shifting, so why should we just limit ourselves? Let's look at quantum computing." Management said, ``That's not a bad idea. Take six months and come back and tell us what you've learned.'' 
\end{quote}
This highlights how preparation can occur through re-framing existing expertise in response to emerging technological directions. Coursework in mathematics and statistics is positioned as enabling this flexibility, providing a foundation that supports engagement with new domains. 

\subsubsection{Trajectory of a government-industry advocate}

One of our interviewees is a government industry advocate at a company that does quantum computing hardware, algorithms and software, and consulting. Their entry into QISE careers combines trajectories one, two, and three (T1, T2, and T3). While formal coursework and research experience provided a technical foundation relevant to quantum technologies, entry into the quantum industry was mainly described by T3 and narrated through a sequence of engagement with quantum-related programs and professional roles that gradually expanded the interviewee’s participation in QISE. 
The interviewee described entering the field through a structured program:
\begin{quote}
    I started quantum through the [educational non-profit organization] program. That's how I broke into the field. They matched me with [company name]. I had a software position with [company name].
\end{quote}
 This structured program  provided the interviewee with access to work experience in software. Subsequent internships and professional roles expanded this participation, including the creation of their own startup. The interviewee highlights how this accumulation of experiences contributed to their hiring:
\begin{quote}
     I would say that my work experience contributed a lot to being able to land the role. Having run my own startup and then being at a startup was really attractive to my boss when I first started. [..] The function of building my own startup and then having the three internships before in quantum and the research allowed my boss to feel confident in my abilities and ability to pick up the material that I was going to be learning in my current role.
\end{quote}
These experiences are described as building familiarity with the field and expanding participation over time. While incremental engagement is central for this interviewee, they also referenced academic research training as providing a technical foundation that supported this participation:
\begin{quote}
    I did focus a lot of my research and studies in quantum computing […] I definitely do rely on the fundamentals that I learned and the knowledge that I had gained in order to be able to communicate with the team effectively and translate the information. [...] I completed research experience my junior year. I was at a nonlinear photonics lab. I learned a lot about EDFA fiber amplifiers and 2D materials during my time there, which also felt valuable for my role now.
\end{quote}
Coursework and research are framed as part of the broader preparation that enabled engagement with QISE. Across the narrative, preparation emerged through the accumulation of these experiences that progressively integrated the interviewee into the field. 

\subsubsection{Trajectory of a test and integration engineer}

One of our interviewees is a test and integration engineer at a company that does quantum networking and communication, computing hardware, and algorithms and software. Their entry into the field aligns with trajectory four (T4), in which entry into the quantum industry is narrated as enabled through relational connections. Although the interviewee describes having relevant technical skills, the decisive moment in their trajectory is a personal connection:
\begin{quote}
    Then I had a friend of mine who was at [company name]. He said, ``Hey, there's this company, they are a startup." At first, I was nervous because I don't see the roadmap being advocated for largely, and you hear a lot of questions about is quantum real? Is it real in the next three decades? He had a meeting with me, and we talked about the technology, what it can do, what exists today, where they plan on being, and with his own eyes, has he seen the system operating and doing quantum operations and solving customer problems. They were solving customer problems. It was a small scope, and I decided like, ``Fine, I will take the leap into this startup and do my best, and hopefully it'll be exciting."
\end{quote}
This participant's access to the opportunity, confidence in the company, and eventual transition into the test and integration position are mediated through a trusted connection already part of the quantum industry. In this case, technical readiness is necessary, but not sufficient. This participant's experience illustrates how access to quantum industry roles is shaped  not only by technical expertise, but also by the strength and influence of professional networks that facilitate and support entry into a quantum career.

\section{Discussion}\label{Discussion}
\subsection{Synthesis}

To answer our research question, we investigated how quantum industry professionals characterize their preparation for participation in the quantum industry. We found that interview participants framed entry into the quantum industry through multiple co-occurring trajectories: continuity of research practice (T1), transfer and reframing of prior expertise (T2), incremental engagement with quantum-related activities (T3), and network-enabled entry (T4). These trajectories, summarized on the right-hand side of Figure \ref{fig:trajectories_co-occurrence}, illustrate how preparation for participation in the quantum industry is better understood as an assembled and socially mediated process rather than a process of simply acquiring knowledge and skills.

A key pattern across interview participants’ narratives concerns the role of formal degrees and coursework within this broader process of preparation. While interviewees consistently describe degrees as establishing a necessary technical foundation, they do not typically frame coursework alone as sufficient for entry into the field. Instead, participants more often frame courses as providing conceptual knowledge or disciplinary language that enables engagement. Preparation is enacted through the ways in which knowledge and skills are mobilized within contexts where they could be applied, demonstrated, and recognized.  

These findings extend existing work on quantum education and workforce development by shifting focus from solely what individuals need to know to how they become positioned and recognized as able to contribute within the field. While prior studies have provided insight into what preparation might entail for course and program development \cite{fox2020preparing,hughes2022assessing, greinert2023future,  greinert2024advancing, Pina2025QuantumWorkforce,  el2025industry, oliver2025education,devendrababu2025mapping, el2026experimental}, our results highlight that possessing relevant KSAs is only one component of entering the field. Across our interviews, participants emphasized not just what they know, but how combinations of varied experiences and interactions with their networks enabled entry into the quantum industry.

Trajectory one (T1) reflects continuity of practice between academic research and industry work. This aligns closely with the situated learning perspective, in which learning is understood as participation in authentic practice rather than simply acquiring KSAs \cite{lave1991situated}. Research training in academia, including doctoral research and undergraduate research, can function as a context of participation in practices that are recognizable as relevant to quantum industry work, which allows individuals to transition between academia and industry, without a fundamental shift in practice. This was also shown in other PER studies, where becoming a physicist is conceptualized as increasing participation in authentic disciplinary practices \cite{irving2015becoming, werth2022impacts}. Our findings contextualize this idea in QISE by showing that academic graduate research can already constitute quantum industry relevant participation. 

Trajectory two (T2) highlights preparation through the reframing of expertise developed in adjacent domains in previous work experiences. This is consistent with the broader STEM career development literature that demonstrates how career pathways often involve movement across domains and require individuals to transfer their expertise to new contexts \cite{ganapati2021professional,edwards2023mapping}. Nevertheless, our findings highlight how that transferability requires individuals to actively construct relevance by reframing their prior work expertise in relation to quantum applications. 

Trajectory three (T3) characterizes preparation as a process of increasing participation in quantum-related activities over time. This trajectory aligns with extensive research on experiential learning in STEM, which shows that internships, undergraduate research, summer schools, and informal learning environments play a central role in shaping career trajectories \cite{rosa2016educational, rethman2021impact, zohrabi2022impact}. Our findings build on this literature by showing how such experiences function within QISE as mechanisms of moving from peripheral exposure toward more central participation in the field. 

Trajectory four (T4) underscores the role of networks in mediating access to the quantum industry. Although participants' technical preparation is key, entry into the field can depend on recognition by others already within it.  This finding is consistent with broader career development research showing that formal and informal networks play a critical role in accessing opportunities \cite{faizan2025importance, zwolak2018informalnetworks, heffernan2021academic}. Our results contextualize this insight for the quantum industry where emerging career pathways may make relational mechanisms for accessing QISE roles particularly salient.  

The co-occurrence of these trajectories in participants' narratives as seen in Figure \ref{fig:trajectories_co-occurrence} shows how preparation emerges through the combination of these experiences rather than through a single pathway. Hence, our findings broaden the focus in quantum education and workforce development by illustrating how entry into the quantum industry depends on how individuals connect experiences across contexts, interpret their relevance, and gain access to opportunities where their knowledge and skills can be demonstrated and validated. This perspective is particularly important for an emerging and interdisciplinary field such as QISE, where roles and career pathways are still evolving.

\subsection{Implications}

Our results have several implications for how preparation for quantum industry careers is conceptualized and supported. 
Our findings show that formal coursework and experiential learning play interrelated roles in preparing students for quantum industry careers. Coursework can help students build conceptual understanding and fluency with the foundational knowledge of the field, while experiential learning opportunities are helpful in applying knowledge and accessing professional communities connected to QISE.

For students, combining multiple forms of engagement may strengthen their ability to enter the quantum industry. In addition to formal coursework, undergraduate students may benefit from seeking opportunities that allow them to apply their knowledge and skills such as undergraduate research experiences, internships, participation in quantum focused programs offered by various organizations (e.g., Qubit by Qubit, which provides training to the future workforce in emerging technologies \cite{QubitByQubit}), and engagement in quantum-focused student organizations (e.g., Quantum Scholars, a program that supports the professional development of undergraduates interested in QISE careers \cite{bennett2024investigatingQuantumScholars}).  Graduate students may benefit from complementing research specialization with broader engagement with the quantum ecosystem, such as participation in industry internships and summer schools. For undergraduate and graduate students alike, building a portfolio of experiences may support entry into QISE by making their skills applicable, visible, and recognizable across multiple contexts.

For educators, our findings highlight the importance of viewing coursework as part of a broader preparation ecosystem. While courses play a critical role in developing foundational knowledge, they can be more impactful when connected to opportunities for application of that course content, such as research experiences, interdisciplinary projects, and extracurricular or professional development opportunities relevant to quantum technologies. Designing curricula that explicitly connect conceptual learning to authentic practices in quantum (e.g.,  industry-coupled project-based quantum capstone course \cite{oliver2025capstone}) may better support students in translating knowledge into participation.

For quantum industry professionals, our results show the importance of experiential learning opportunities and professional networks in shaping access to the field. As such, quantum industry professionals have the opportunity to create and expand pathways that allow students to gain these relevant experiences to enable entry. The continued investment by current industry professionals in university partnerships, participation in conferences, and the development and expansion of internship programs that are accessible to all degree levels can help students access networks and opportunities offered by industry.  However, reliance on personal networks for hiring may limit connections between employers and qualified and ambitious candidates graduating from less-resourced institutions who may not have existing industry connections. Therefore, the quantum industry will benefit from continued efforts  to create structured, transparent, and broadly accessible entry points in order to broaden participation to all students interested in quantum careers.

Our findings lay the foundation for further exploration of how relevant experiences can be made visible to employers during recruitment and application processes. Future work could potentially investigate the different ways students can gain access to opportunities and be recognized for having relevant KSAs by employers in order to address some of the challenges students perceive in obtaining jobs in the quantum industry \cite{oliver_us_students_survey}.

In conclusion, preparation for participation in the quantum industry consists of a dynamic combination of educational, experiential, and relational factors. Understanding these trajectories provides a more complete picture of how individuals enter this emerging field by broadening the focus beyond knowledge and skills. Our findings inform ongoing efforts in quantum education and workforce development, particularly as education and training structures that support participation are actively being designed and refined.

\begin{acknowledgments}
We would like to thank the quantum industry professionals who participated in our interviews. This work is based on work supported by the National Science Foundation under Grant Nos. PHY-2333073 and PHY-2333074. This work is also based on work supported by the Army Research Office and was accomplished under Award Number: W911NF-24-1-0132. 
\end{acknowledgments}



\bibliography{apssamp}
\end{document}